\documentclass[12pt]{iopart}
\usepackage{iopams}
\expandafter\let\csname equation*\endcsname\relax
\expandafter\let\csname endequation*\endcsname\relax
\usepackage[fleqn]{amsmath}
\usepackage{amssymb}
\usepackage{amsbsy}
\usepackage{color}
\usepackage{graphicx}

\newcommand\apjl{The Astrophysical Journal Letters}
\newcommand\apj{The Astrophysical Journal}
\newcommand\mnras{Monthly Notices of the Royal Astronomical Society}
\newcommand\prd{Physical Review D}

\newcommand{\iu}{\mathrm{i}\mkern1mu}

\begin{document}

\title{Fundamentals of the orbit and response for TianQin}

\author{Xin-Chun Hu$^1$, Xiao-Hong Li$^1$, Yan Wang$^1$, Wen-Fan Feng$^1$, Ming-Yue Zhou$^1$, 
Yi-Ming Hu$^{2,3}$, Shou-Cun Hu$^{4,5}$, Jian-Wei Mei$^{2,3}$ and Cheng-Gang Shao$^1$}

\address{$^1$ MOE Key Laboratory of Fundamental Physical Quantities Measurements, 
Hubei Key Laboratory of Gravitation and Quantum Physics, 
School of Physics, Huazhong University of Science and Technology,
1037 Luoyu Road, Wuhan 430074, China}

\address{$^2$ TianQin Research Center for Gravitational Physics, 
Sun Yat-Sen University, Zhuhai 519082, China}

\address{$^3$ School of Physics and Astronomy, Sun Yat-Sen University,
Zhuhai 519082, China}

\address{$^4$ Key Laboratory of Planetary Sciences, Purple Mountain Observatory, 
Chinese Academy of Sciences, 8 Yuanhua Road, Nanjing 210008, China} 

\address{$^5$ University of Chinese Academy of Sciences, Beijing 100049, China}

\ead{ywang12@hust.edu.cn} \vspace{10pt}
\begin{indented}
\item[] March 5, 2018
\end{indented}

\begin{abstract}

TianQin is a space-based laser interferometric gravitational wave detector aimed at 
detecting gravitational waves at low frequencies (0.1 mHz -- 1 Hz). It is formed by three 
identical drag-free spacecrafts in an equilateral triangular constellation orbiting around 
the Earth. The distance between each pair of spacecrafts is approximately 
$1.7 \times 10^{5} ~\rm{km}$. 
The spacecrafts are interconnected by infrared laser beams forming up to three 
Michelson-type interferometers. The detailed mission design and the study of science 
objectives for the TianQin project depend crucially on the orbit and the response of the 
detector. In this paper, we provide the analytic expressions for the coordinates of the 
orbit for each spacecraft in the heliocentric-ecliptic coordinate system to the leading orders. 
This enables a sufficiently accurate study of  science objectives and data analysis, and serves as 
a first step to further orbit design and optimization.  We calculate the response of 
a single Michelson detector to plane gravitational waves in arbitrary waveform 
which is valid in the full 
range of the sensitive frequencies. It is then used to generate the more realistic 
sensitivity curve of TianQin. We apply this model on a reference white-dwarf binary 
as a proof of principle. 

\end{abstract}

\vspace{2pc} \noindent{\it Keywords}: gravitational waves,
space-borne detector, TianQin

\section{Introduction}\label{sec:intro}

The first direct detection of gravitational waves (GWs) from a pair of 
stellar-mass black holes (GW150914) has been made by the two 
advanced detectors of the Laser Interferometer Gravitational-Wave 
Observatory (LIGO) \cite{2016PhRvL.116f1102A}. This event opens 
the era of observational GW astronomy. 
Subsequently, several more GW events of stellar-mass black hole 
binaries have been detected 
\cite{2016PhRvL.116x1103A, 2017PhRvL.119n1101A, 
2017PhRvL.118v1101A,2017ApJ...851L..35A}. 
The first detection of GWs from a pair of neutron stars 
associated with a gamma-ray burst (GW170817/GRB170817A) 
\cite{2017PhRvL.119p1101A} 
has been made jointly by advanced LIGO and 
advanced Virgo \cite{2015CQGra..32b4001A} 
which marks a major breakthrough of the multi-messenger astronomy. 
In the future, KAGRA \cite{2012CQGra..29l4007S} and LIGO-India \cite{2013IJMPD..2241010U}
will also join in the ground-based detector network.
Developments for the upgraded ground-based detectors, namely Advanced LIGO plus \cite{2013PhRvD..88b2002E, 2015PhRvD..91f2005M} and LIGO Voyager, as well as the next generation detectors, such as Einstein Telescope \cite{2011CQGra..28i4013H} and Cosmic Explorer \cite{2017CQGra..34d4001A}, have already been initiated.

Lower frequency windows for the GW astronomy are awaiting to be opened. 
Pulsar timing array (PTA), as a promising experiment sensitive to the very 
low frequency ($10^{-9}-10^{-6}$~Hz) GWs, has been continuously improving 
its sensitivity for stochastic background produced by the incoherent 
superposition of GWs from a large ensemble of supermassive 
black hole binaries 
\cite{2015Sci...349.1522S, 2015MNRAS.453.2576L, 2016ApJ...821...13A} 
and continuous waves from resolvable individual sources 
\cite{2014ApJ...794..141A,  2014MNRAS.444.3709Z,  2016MNRAS.455.1665B}. 
Prospects for GW astronomy with next generation large-scale 
PTAs based on FAST \cite{2014arXiv1407.0435H} and 
SKA \cite{2017PhRvL.118o1104W} have been 
investigated  \cite{2017arXiv171104435W}.

Due to seismic noise and gravity gradient noise, it is very difficult 
to detect GWs with frequency lower than $\sim 10$ Hz 
by the ground-based detectors. It is a natural choice to deploy 
the laser interferometers in space \cite{1985ESASP.226..157F}. 
There have been long-term studies and plots for the space-borne 
detectors to explore the low frequency regime (0.1 mHz -- 1 Hz), 
where there is a very rich source of GWs \cite{Sathyaprakash2009}. 
LISA/eLISA \cite{2017arXiv170200786A, 2013arXiv1305.5720C} is 
the most well studied one, especially with the successful demonstration 
of gravity reference system 
and space laser interferometry by the 
LISA Pathfinder \cite{PhysRevLett.116.231101}. The other space 
laser interferometric detectors, such as TianQin \cite{2016CQGra..33c5010L}, 
DECIGO \cite{2009JPhCS.154a2040S}, g-LISA/GEOGRAWI 
\cite{Tinto2011Searching,Tinto2015gLISA}, 
ASTROD-GW \cite{2002IJMPD..11..947N, Ni2009ASTROD}, 
TAIJI (ALIA descoped) \cite{2015JPhCS.610a2011G}, 
BBO \cite{Cutler2006BBO,Crowder2005Beyond}, 
OMEGA \cite{Hiscock1997OMEGA} 
and LAGRANE \cite{Conklin2011LAGRANGE}, are currently 
under various stages of research and development. A thorough 
overview of space-borne GW detectors can be found 
in \cite{2016IJMPD..2530001N}.

During the initial study phase of the TianQin project, 
not all issues/complexities pertinent to the detector are fully 
recognized or anticipated.  Work is now in progress to build 
prototype and end-to-end model to bring most of the issues 
into surface in order to avoid possible mistakes in the full 
development. The end-to-end model will utilize realistic simulations 
based on the feasible technologies to study the impact of 
different subsystems of the detector on the final high-precision 
measurements \cite{2003CQGra..20S.255M}. In principle, 
the elements of an end-to-end model will include the orbit of 
the spacecrafts \cite{2005CQGra..22..481D}, interplanetary 
and relativistic effects \cite{2005PhRvD..72l2003C} on the 
optical links, various noises in a comprehensive noise budget, 
and the implementation of time delay interferometry (TDI) to 
reduce the laser phase noise and optical bench 
noise \cite{2005LRR.....8....4T}. 
This model will aid in the design and optimization of the 
TianQin detector, so that it can meet the science requirements 
as launched.

The current work, as our first step toward building the end-to-end model, 
is to provide the mathematical formalism to represent the coordinates of 
the fiducial orbit for the TianQin's spacecraft. 
In the preliminary proposal \cite{2016CQGra..33c5010L}, 
the TianQin detector is composed of three drag-free spacecrafts in an 
equilateral triangular constellation orbiting around the Earth (see Fig.~\ref{fig:1}). 
The guiding center of the constellation coincides with the geocenter.
The geocentric distance of each spacecraft is $1.0 \times 10^{5} ~\rm{km}$, 
which makes the distance between each pair of spacecrafts be 
$1.7 \times 10^{5}~\rm{km}$. The period of the nearly circular 
Keplerian orbit of the spacecraft around the Earth is 
approximately $ 3.65$ day.  The spacecrafts 
are interconnected by infrared laser beams and form up to 
three (not independent) Michelson-type interferometers. The normal 
of the detector plane formed by the constellation points toward 
the tentative reference source 
RX J0806.3+1527 (also known as HM Cancri or HM Cnc), which is 
a candidate ultracompact white-dwarf binary 
in the Galaxy \cite{2041-8205-711-2-L138}.

The time-varying spacecraft coordinates will be subsequently used 
in the forward modeling of the detector's response to the incident 
GWs. The calculation will be realized by the science 
data simulator module (e.g., LISACode \cite{2008PhRvD..77b3002P} 
for LISA) in the end-to-end model, which is essentially used to 
convert the tensor GW perturbation to the scalar strain 
measurable by the detector.  Noise components can also be added here. 
This simulator is crucial for studying the science objectives that 
can be enabled by TianQin \cite{doi:10.1093/nsr/nwx115}  
and testing the science data analysis 
techniques and pipelines of various GW sources. 

The response of a space-based laser interferometer to GWs is not a straightforward 
extension of its ground-based counterparts. For the latter, geodesic 
deviation is used to calculate the proper distance change between two 
test masses. This treatment is valid only when the detector arm length $L$ is shorter 
than the reduced wavelength of the passing GWs (i.e., under long-wavelength 
or low-frequency approximation) \cite{1987thyg.book..330T}. 
The critical wavelength corresponds to the so-called \textit{transfer 
frequency} $f_{\ast}=c/(2\pi L)$.  The sensitive GW frequency band 
is far below $f_{\ast}$ for the ground-based detectors, but not for 
the space-based detectors. 
For frequencies higher than $f_{\ast}$ ($\approx 0.28~\rm{Hz}$ for TianQin), 
the GW effect starts to cancel itself which in turn deteriorates the detector response. 
To obtain the time-varying detector response that is applicable 
in the whole frequency band, in principle one needs to integrate along the 
null geodesics of the photon that connect free-fall test 
masses \cite{2003PhRvD..67b2001C}. 
However, this approach is computationally expensive. 
Instead, we adopt the \textit{rigid adiabatic} approximation of the full 
response which is proven to have the high fidelity that meets the 
requirements of our subsequent work while significantly reduces 
the computational cost \cite{2004PhRvD..69h2003R}.

The rest of the paper is organized as follows. 
In Section~\ref{sec:orbit}, we present the coordinates for the orbit 
of TianQin's spacecraft, which are then used in the calculation of 
the detector response to gravitational waves and detector's 
sensitivity curve in Section~\ref{sec:response}.
An example signal from the reference white-dwarf binary is given 
in Section~\ref{sec:simulation}. 
We conclude the main part of the paper in Section~\ref{sec:sum}. 
Some calculation details have been relegated to the Appendix 
in order to allow 
the main ideas of the paper to be as clear as possible.

\section{Spacecraft Orbit}\label{sec:orbit}

\begin{figure}[!h]
\centering
\includegraphics[width=0.8\textwidth]{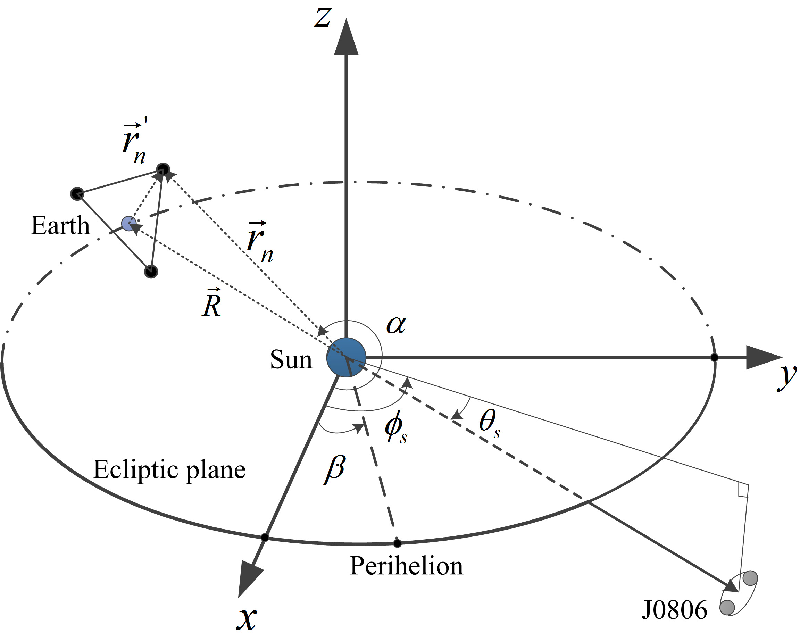}
\caption{Schematic of the TianQin spacecraft orbit in the 
heliocentric-ecliptic coordinate system. The ecliptic plane is 
spanned by $x$ and $y$ axes. $x$ axis points toward the 
direction of the vernal equinox. $\beta$ is the longitude of the 
perihelion. $\phi_s$ and $\theta_s$ are the longitude and 
latitude of the reference source RX J0806.3+1527, 
in which direction the normal of the detector plane formed 
by the three spacecrafts is placed. $\alpha$ is the longitude 
of the guiding center of the spacecraft constellation. 
$\mathbf{r}_n(t)$ and $\mathbf{r}'_n(t)$ are the position vectors 
of the $n$-th spacecraft relative to the heliocenter and the 
guiding center, respectively. $\mathbf{R}(t)$ is the position 
vector of the guiding center measured from the heliocenter. 
} \label{fig:1}
\end{figure}

For a long-term space mission, it is natural to represent the 
coordinates of the spacecraft orbit in the heliocentric-ecliptic 
coordinate system shown in Fig.~\ref{fig:1}. The center of the 
coordinate system locates at the solar system heliocenter, 
the ${x}$ axis points in the direction of the vernal equinox, 
the ${z}$ axis is parallel to the normal of the ecliptic plane, 
and the ${y}$ axis is placed in the ecliptic plane to complete 
the right-handed coordinate system. The position vector of 
the spacecraft $\mathbf{r}_n(t)$ can be decomposed into the 
position vector relative to the guiding center $\mathbf{r}'_n(t)$ 
and the position vector of the guiding center relative to the 
heliocenter $\mathbf{R}(t)$ as follows 
\begin{equation}\label{eq:sum}
\mathbf{r}_n(t)=\mathbf{r}'_n(t)+\mathbf{R}(t)+\boldsymbol{\varepsilon}_n(t)  \,,   \\
\end{equation}
where $n \in\lbrace1,2,3\rbrace$ is the index of the spacecraft. 
$\mathbf{r}'_n(t)=(x'_n(t),y'_n(t),z'_n(t))$ and 
$\mathbf{R}(t)=(X(t),Y(t),Z(t))$ trace the nearly circular Keplerian 
orbit around the Earth and the Sun, respectively. 
$\boldsymbol{\varepsilon}_n(t)$  is the correction to the Keplerian 
orbits which takes into account of the perturbations from 
high order moments of the Earth, the Sun, the moon, 
planets (mostly Jupiter), large asteroids, etc. 
We give the components of the $n$-th spacecraft's position vector 
$\mathbf{r}_n(t)=(x_n(t),y_n(t),z_n(t))$ as a function of time 
in the heliocentric-ecliptic coordinate system below. 
The detailed derivation is relegated to \ref{sec:coords}. 
\begin{eqnarray}\label{eq:heliellip}
  x_n(t)&&=R_1\big(\cos\phi_s\sin\theta_s\sin(\alpha_n-\beta')+\cos(\alpha_n-\beta')\sin\phi_s\big)+R_1{e_1}\Big[\frac{1}{2}\big(\cos2(\alpha_n-\beta')-3\big)\sin\phi_s\nonumber\\
  &&+\cos(\alpha_n-\beta')\cos\phi_s\sin\theta_s\sin(\alpha_n-\beta')\Big]+\frac{e^2_1}{4}R_1\sin(\alpha_n-\beta')\Big[\big(3\cos2(\alpha_n-\beta')-1\big)\nonumber\\
  &&\times\cos\phi_s\sin\theta_s-6\cos(\alpha_n-\beta')\sin(\alpha_n-\beta')\sin\phi_s\Big]+R\cos(\alpha-\beta)+\frac{Re}{2}\big(\cos2(\alpha-\beta)-3\big)\nonumber\\
  &&-\frac{3R{e^2}}{2}\cos(\alpha-\beta)\sin^2(\alpha-\beta)  \,,  \nonumber\\
  y_n(t)&&=R_1\big(\sin\phi_s\sin\theta_s\sin(\alpha_n-\beta')-\cos(\alpha_n-\beta')\cos\phi_s\big)-R_1{e_1}\Big[\frac{1}{2}\big(\cos2(\alpha_n-\beta')-3\big)\cos\phi_s\nonumber\\
  &&-\cos(\alpha_n-\beta')\sin\phi_s\sin\theta_s\sin(\alpha_n-\beta')\Big]+\frac{e^2_1}{4}R_1\sin(\alpha_n-\beta')\Big[\big(3\cos2(\alpha_n-\beta')-1\big)\nonumber\\
  &&\times\sin\phi_s\sin\theta_s+6\cos(\alpha_n-\beta')\sin(\alpha_n-\beta')\cos\phi_s\Big]+R\sin(\alpha-\beta)+\frac{Re}{2}\sin2(\alpha-\beta)\nonumber\\
  &&+\frac{R{e^2}}{4}\big(3\cos2(\alpha-\beta)-1\big)\sin(\alpha-\beta)   \,,  \nonumber\\
  z_n(t)&&=-R_1\sin(\alpha_n-\beta')\cos\theta_s-R_1{e_1}\cos(\alpha_n-\beta')\sin(\alpha_n-\beta')\cos\theta_s\nonumber\\
  &&-\frac{1}{4}{e^2_1}R_1\big(3\cos2(\alpha_n-\beta')-1\big)\sin(\alpha_n-\beta')\cos\theta_s   \,, 
\end{eqnarray}
where $R_1=1.0\times 10^{5} ~\rm{km}$ and ${e_1}$ are the semi-major 
axis and the eccentricity of the spacecraft orbit around the Earth; $R=1$ AU 
and $e=0.0167$ are the semi-major axis and the eccentricity of the geocenter 
orbit around the Sun. 
The normal of the detector plane formed by the spacecraft 
constellation points along the direction of 
$\left( \theta_{s}=- 4.7^{\circ}, \phi_{s}=120.5^{\circ} \right)$ 
which is the sky position of the reference source RX J0806.3+1527 
in the ecliptic coordinates. 
Here $\alpha (t) = 2\pi f_{\rm{m}}t + \kappa_0$ is the mean ecliptic longitude 
of the geocenter in the heliocentric-ecliptic coordinate system, 
$f_{\rm{m}}=1/\rm{(one~sidereal~year)} = 3.14 \times 10^{-8}~\rm{Hz}$ 
is the modulation frequency due to the orbital motion around the Sun, 
and $\kappa_0$ is the mean ecliptic longitude measured from 
the vernal equinox at $t=0$. $\beta$ is the longitude of the perihelion. 
To obtain Eq.~\ref{eq:heliellip}, 
we have introduced the detector coordinate system 
(see details in \ref{sec:coords}) 
in which $\alpha_n$ is defined as the orbit phase of 
the $n$-th spacecraft in the detector 
plane and $\beta'$ is the angle measured from the 
$\tilde{x}$ axis in Fig.~\ref{fig:2} to the 
perigee of the spacecraft. 
As stated in \ref{sec:coords}, we have kept only up to the quadratic 
terms of the eccentricities $e$ and $e_1$ here. 
Including higher order terms is straightforward.

Using Eq.~\ref{eq:heliellip}, one can find that 
the instantaneous distance between the spacecraft $i$ and $j$ is 
\begin{eqnarray}\label{eq:armlength}
 {L_{12}}(t)&&= \sqrt 3 R_1\Big[1 +\frac{e_1}{4}\big(\cos(\kappa-\beta') -\sqrt 3 \sin(\kappa-\beta') \big)\nonumber\\
 &&+\frac{e_1^2}{32}\big( 3\sqrt 3 \sin2(\kappa-\beta')-14 +3\cos2(\kappa-\beta')\big)\Big] + {\rm O}({e^3_1})  \,,  \nonumber\\
 {L_{13}}(t)&&=\sqrt 3 R_1\Big[1 +\frac{e_1}{4}\big(\cos(\kappa-\beta') + \sqrt 3 \sin(\kappa-\beta') \big)\nonumber\\
 &&-\frac{e_1^2}{32}\big( 3\sqrt 3 \sin2(\kappa-\beta')+14 -3\cos2(\kappa-\beta')\big)\Big]+ {\rm O}({e^3_1})  \,,  \nonumber\\
 {L_{23}}(t)&&= \sqrt 3 R_1\Big[1 -\frac{e_1}{2}\cos(\kappa-\beta') -\frac{e_1^2}{16}\big(7+ 3\cos2(\kappa-\beta')\big)\Big] + {\rm O}(\!{e^3_1})  \,,
\end{eqnarray}
where $\kappa (t) = 2\pi f_{\rm{sc}}t + \lambda $, 
$f_{\rm{sc}} \approx 1/(3.65~\rm{day})$ is the modulation 
frequency due to the rotation of the spacecrafts around the 
guiding center, $\lambda$ is the angle between the first $(n=1)$ 
spacecraft and 
$\tilde x$ axis (see Fig.~\ref{fig:2}) at $t=0$. 
We can see from Eq.~\ref{eq:armlength} that to the leading order 
of $e_1$ the distance between two spacecrafts is 
$\sqrt 3 {R_1} \approx 1.7 \times {10^5}  ~\rm{km}$ 
which is the fiducial arm length of TianQin \cite{2016CQGra..33c5010L}. 
Note that a smaller $e_1$ is preferred to reduce the arm length variation 
during the mission. For example, one need to set 
$e_1< 0.05$ when deploying the spacecraft in order to suppress the 
range rate of each arm below $15~\rm{m~s}^{-1}$ that is 
required by space interferometry \cite{2016CQGra..33c5010L}.

Besides neglecting the higher order terms of the eccentricities, 
Eq.~\ref{eq:heliellip} has not yet included the correction term 
$\boldsymbol{\varepsilon}_n(t)$.

\section{Detector Response}\label{sec:response}

\subsection{Description of gravitational waves}\label{subsec:gw}

The plane GWs in arbitrary waveform can be decomposed into 
a spectrum of monochromatic waves with two independent polarizations, 
which can be expressed in the source frame as 
%
\begin{equation}\label{eq:gwsource}
{\bf{h}}(t,\mathbf{r})=\int_{-\infty}^{+\infty} df \exp\left(\iu 2\pi f (t-\hat{\Omega}\cdot \mathbf{r}/c) \right) \sum_{A=+, \times} \tilde{h}_{A}(f){\boldsymbol{\epsilon}}^A(\hat \Omega) \,. 
\end{equation}
Here $\iu$ is the imaginary unit, $\hat \Omega$ is the unit vector pointing along the 
direction of GW propagation and $c$ is the speed of light. 
$\tilde{h}_{A}(f)$ ($A\in \lbrace +, \times \rbrace$) are the Fourier transforms 
of two GW polarizations. ${\boldsymbol{\epsilon}}^A(\hat \Omega)$ 
are the basis tensors, which are connected with the basis tensors 
${\bf{e}}^A(\hat k)$ defined in observer's frame 
%
%
via the polarization angle $\psi$ in 
the standard spin-2 rotation transformation \cite{2009LRR....12....2S} 
\begin{eqnarray}\label{eq:basistrans}
\begin{array}{l}
{\boldsymbol{\epsilon}}^{+}(\hat \Omega)={\bf{e}}^+(\hat k)\cos2\psi - {\bf{e}}^\times(\hat k)\sin2\psi  \,, \\
{\boldsymbol{\epsilon}}^{\times}(\hat \Omega)={\bf{e}}^+(\hat k)\sin2\psi + {\bf{e}}^\times(\hat k)\cos2\psi \,. 
\end{array}
\end{eqnarray}
Here  $\hat \Omega=-\hat k$. 
${\bf{e}}^A(\hat k)$ can be formed by the unit 
vectors $(\hat p,\hat q)$ 
%
\begin{eqnarray}\label{eq:basistensor}
\begin{array}{l}
\mathbf {{e^+}}  = \hat p \otimes \hat p - \hat q \otimes \hat q  \,, \\
\mathbf {{e^\times }} = \hat p \otimes \hat q + \hat q \otimes \hat p \,, 
\end{array}
\end{eqnarray}
where $\otimes$ is the tensor product, 
$\hat p$ and $\hat q$ reside in observer's plane of sky paralleling to 
the tangent vectors of $\theta$ and $\phi$ in the heliocentric ecliptic 
coordinate system, respectively. 
$(\hat p,\hat q,\hat k)$ are orthonormal basis vectors which can be 
expressed as 
\begin{eqnarray}\label{eq:basisvector}
\begin{array}{l}
\hat p = \left(\cos \theta \cos \phi ,\cos\theta \sin\phi , - \sin \theta \right) \,, \\
\hat q = \left(\sin\phi , - \cos \phi ,0 \right)  \,,  \\
\hat k = \left(\sin\theta \cos\phi ,\sin\theta \sin\phi ,\cos\theta \right)  \,.
\end{array}
\end{eqnarray}

\subsection{Detector response}\label{subsec:response}

For space-borne GW interferometer, the detector response, 
i.e. strain, is defined as the time variation of the 
differential optical path length of 
the two arms divided by the optical path length of one arm. 
Here, we adopt the rigid adiabatic 
approximation\footnote{The term ``adiabatic" indicates that 
we approximate the continuous motion of the spacecraft 
constellation as a series of quasi-stationary state during the 
photon traveling time between two spacecrafts (equivalently 
taking stroboscopic images of the constellation).} \cite{2004PhRvD..69h2003R} 
to get the closed form of the strain output for 
the Michelson interferometer centered on spacecraft $i$ 
%
\begin{equation}\label{eq:strain} 
h_i(t)=\int_{-\infty}^{+\infty} df \exp\left(\iu 2\pi f (t-\hat{\Omega}\cdot \mathbf{r}_i/c) \right) \sum_{A=+, \times} \tilde{h}_{A}(f) F_i^{A}(t;\hat k,f) \,. 
\end{equation}
Here ${F_i^A(t;\hat k,f)}$  
are called the \textit{antenna pattern functions} 
which are generally functions of source direction $\hat k$ 
and GW frequency $f$. They are defined by \cite{2009LRR....12....2S} 
\begin{equation}\label{eq:apf} 
F_i^{A}(t;\hat k,f) \equiv {\bf{D}}_i(t;\hat k,f):{\boldsymbol{\epsilon}}^A(\hat \Omega)  \,,
\end{equation}
where the colon represents the double contraction of two 
rank-2 tensors, i.e., ${\bf{A}}:{\bf{B}}=A_{ij}B^{ij}$. 
${\bf{D}}_i$ is the \textit{detector tensor} of the 
interferometer centered on spacecraft $i$ 
\begin{equation}\label{eq:Dtensor}
{\bf{D}}_i(t;\hat k,f)=\frac{1}{2}\Big[{\hat r_{ij}(t)}\otimes {\hat
r_{ij}(t)}{\rm{{\cal T}}}({\hat r_{ij}(t)},{\hat k,f)}-{\hat
r_{ik}(t)}\otimes {\hat r_{ik}(t)}{\rm{{\cal T}}}({\hat
r_{ik}(t)},{\hat k,f)}\Big]  \,,
\end{equation}
where $i,j,k \in\lbrace 1,2,3 \rbrace$ and $i \neq j \neq k$. 
$\hat r_{ij}(t)$ is the unit vector points from spacecraft $i$ 
to spacecraft $j$. 
${\cal T}$ is the \textit{transfer function} 
which can be written as \cite{2003AdSpR..32.1277H} 
\begin{eqnarray}\label{eq:transferfunc}
{\cal T}(\hat r_{ij}(t),\hat
k,f)&&=\frac{1}{2}\big[\sin\!{\mathop{\rm c}\nolimits}\big(\frac{f}{2f_*}(1+\hat k \cdot {\hat r_{ij}}\left( t \right))\big)\exp\big(-\iu \frac{f}{2f_*}(3-\hat k \cdot {\hat r_{ij}}\left( t \right) )\big)\nonumber\\
&&+\sin\!{\mathop{\rm c}\nolimits}\big(\frac{f}{2f_*}(1-\hat k
\cdot {\hat r_{ij}}\left( t\right))\big)\exp\big(-\iu \frac{f}{2f_*}(1-\hat k \cdot {\hat r_{ij}}\left( t \right))\big)\big]  \,,
\end{eqnarray}
where $f_{*} = c /(2\pi L_{ij})$ is the transfer frequency. 
As we shall see in Sec.~\ref{subsec:sensitivity}, 
in the frequency region that $f>f_{*}$ the detector response 
starts to oscillate and decay that partly cancel the GW effect 
in accord with the characteristics of $\cal T$. 
${\cal T}$ approaches unity if $f< f_{*}$, 
such that the detector tensor 
becomes dependent only on the geometric configuration of 
the interferometer.  Note that $f_{*}$ is not a constant due to the 
variation of the arm length induced by the breathing and flexing 
of the constellation. 
%

We define a surrogate variable 
$\xi_i^A(t) = {\bf{D}}^A_i(t;\hat k,f):{\bf{e}}^A(\hat k)$. 
For $i=1$ 
\begin{eqnarray}\label{eq:projfunc}
 \xi_1^+(t;\theta,\phi)&&=\frac{\sqrt3}{32}\big[4\cos2(\kappa-\beta')\big((3+\cos2\theta)\sin\theta_s\sin2(\phi\!-\!\phi_s)+2\sin(\phi\!-\!\phi_s)\sin2\theta\cos\theta_s\big)\nonumber\\
 &&-\sin2(\kappa-\beta')\big(3+\cos2(\phi\!-\!\phi_s)\big(9+\cos2\theta(3-\cos2\theta_s)\big)\!+6\cos2\theta_s{\sin}^2(\phi\!-\!\phi_s)  \nonumber\\
 &&-6\cos2\theta{\cos}^2\theta_s+4\cos(\phi\!-\!\phi_s)\sin2\theta\sin2\theta_s\big)\big]  \,,\\
 \xi_1^\times(t;\theta,\phi)\!&&=\frac{\sqrt3}{8}\big[-4\cos2(\kappa-\beta')\big(\cos2(\phi\!-\!\phi_s)\cos\theta\sin\theta_s+\cos(\phi\!-\!\phi_s)\sin\theta\cos\theta_s\big)\nonumber\\
 &&+\sin2(\kappa-\beta')\big(\cos\theta(3-\cos2\theta_s)\sin2(\phi_s\!-\!\phi)+2\sin(\phi_s\!-\!\phi)\sin\theta\sin2\theta_s\big)\big]  \,. \nonumber
\end{eqnarray}
The explicit expressions for $\xi_2^{+,\times}$ and $\xi_3^{+,\times}$ 
can be similarly obtained. Note that for simplicity we have set 
$e_1=0$ in Eq.~\ref{eq:projfunc}, considering that $\xi_i^{+,\times}$ 
in expansion of $e_1$ are straightforward to derive but 
too lengthy to be shown here. 
Combining Eq.~\ref{eq:basistrans}, Eq.~\ref{eq:apf} and $\xi_i^{+,\times}$, 
we can get 
\begin{eqnarray}\label{eq:apf2}
F_i^+(t;\theta,\phi,\psi ) = \cos 2\psi \, {\xi_i^{ + }}(t;\theta,\phi) - \sin 2\psi \, {\xi_i^{ \times }} (t;\theta,\phi)  \,,  \nonumber\\
F_i^\times(t;\theta,\phi,\psi) =\sin 2\psi \, {\xi_i^{ + }}(t;\theta,\phi) + \cos 2\psi \, {\xi_i^{ \times }} (t;\theta,\phi)  \,. 
\end{eqnarray}

\subsection{Sensitivity curve}\label{subsec:sensitivity}

The noise budget for the key components of the TianQin detector has been 
given in \cite{2016CQGra..33c5010L}. The main ones are the position 
noise (optical path noise) rooted from the photon shot noise at 
the measurement of laser phase 
by phasemeter and the acceleration noise 
rooted from the initial sensor in the disturbance reduction system. 
The preliminary goal for the amplitude spectral density (ASD) of these two 
noise sources are $\sqrt{S_x} = 1 ~\text{pm}/\text{Hz}^{1/2}$ 
for each position measurement and 
$\sqrt{S_a}=10^{-15}~\text{m~s}^{-2}/\text{Hz}^{1/2}$ 
for each interaction with inertial sensor, respectively.  These two noises can be 
essentially regarded as white in the interested frequency range, 
except that the inherent $1/f$ noise will emerge from the inertia 
sensor for $f<10^{-4} ~\rm{Hz}$.

The position noise from each phasemeter measurement is 
uncorrelated. Thus simply summing over the power spectral 
density (PSD) contribution from 
each phasemeter in the optical path gives the total PSD. 
The equivalent strain noise ASD $\sqrt{S_n^{x}}$ 
can be obtained through dividing the total position noise 
ASD by the optical path length of $2L$ on one arm, which gives 
\begin{equation}\label{eq:xnoise}
S_n^{x}= \left( \frac{\sqrt{4S_x}}{2L} \right)^2  =  \frac{S_x}{L^2}   \,. 
\end{equation}
The factor of four in the numerator is due to the fact that 
there are two phasemeter measurements 
on each arm (one at the transmitting spacecraft, 
one at the responding spacecraft) which add up to four 
measurements for the Michelson interferometer.  

Similarly one can calculate the equivalent strain noise due to acceleration noise. 
Note that the acceleration noise acts coherently on the incoming and 
outgoing laser, which gives a combined $2\sqrt{S_a}$ for each inertial 
sensor on board of a spacecraft. There are four contributions for 
an interferometer. Dividing the total acceleration noise ASD 
by $(2\pi f)^2$ gives the equivalent position noise ASD, further 
dividing it by $2L$ gives the equivalent strain noise ASD, 
\begin{equation}\label{eq:anoise}
S_n^{a}= \left( \frac{\sqrt{16 S_a}}{(2\pi f)^2 \cdot 2L} \right)^2 =  \frac{4 S_a}{(2\pi f)^4 L^2}  \,. 
\end{equation}

\begin{figure}
\centering
\includegraphics[width=0.8\textwidth]{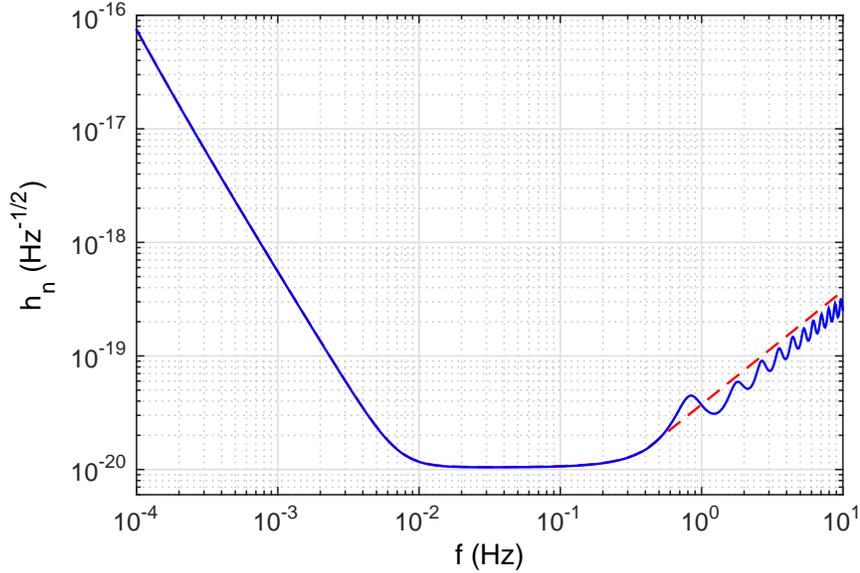} \\
\caption{Sensitivity curve of TianQin. The blue solid line is the numerical evaluation 
of Eq.~\ref{eq:strainApsd} over the interested frequency range. 
The red dash line is obtained from an analytic approximation. The two lines 
overlap for $f<f_{*} \approx 0.28~\rm{Hz}$. }
\label{fig:sensitivitycurve}
\end{figure}

Following \cite{2016CQGra..33c5010L} we have the ASD  
of the effective strain noise (in unit of $1/\sqrt{\rm{Hz}}$)
\begin{equation}\label{eq:strainApsd}
h_n(f)=\frac{1}{\sqrt{R(f)}}\bigg[ S_n^{x}   +S_n^{a}\left(1+\frac{10^{-4}~ \mathrm{Hz}}{f}\right)\bigg]^{\frac{1}{2}}  \,, 
\end{equation}
here the multiplier associated with $S_n^{a}$ is owning to 
the $1/f$ noise of the inertial sensor. 
The function $R(f)$ is defined as follows \cite{0264-9381-18-17-308},
\begin{equation}\label{eq:Rfunc}
R(f)= \int\frac{d\hat{k}}{4\pi}\sum\limits_A {F^ A }\left( t;{\hat k,f } \right){F^ A }\left( t;{\hat k,f } \right)^*  \,, 
\end{equation}
which is an all-sky average for the sum of squared amplitudes of the 
antenna pattern functions. Note that the polarization angle $\psi$ 
has been self canceled in the integrand. 
Based on Eq.~\ref{eq:xnoise}--\ref{eq:Rfunc}, the sensitivity curve, i.e., 
$h_n$ of TianQin 
is calculated and presented in Fig.~\ref{fig:sensitivitycurve}. 
The red dash line represents an analytic approximation of $R(f)$ 
\cite{0264-9381-18-17-308} 
which is in a good agreement with the approximated sensitivity curve 
shown in Fig. 3 of \cite{2016CQGra..33c5010L}. 
Under this approximation, the wiggles in $f>f_{\ast}$ 
originated from the sinc function in $\cal T$  have been smoothed out. 
As we can see, the frequency of the sensitivity curve can be divided into three regions: 
for $f<10^{-2} ~\rm{Hz}$ the curve is determined by the acceleration noise, 
such that the sensitivity is worsened as $f^{-2}$ (according to Eq.~\ref{eq:anoise}); 
for $10^{-2} ~{\rm{Hz}}<f<f_{\ast}\approx 0.28~\rm{Hz}$, i.e., the flat bottom, 
the curve is determined by the frequency independent position noise; 
for $f>f_{\ast}$ the curve is determined by $R(f)$. 

\section{Signal Simulation}\label{sec:simulation}

Taking the reference source, J0806.3+1527, as an example, we 
simulate of the strain output from a single Michelson interferometer. 
The quadrupole formula provide 
the lowest-order post-Newtonian GW waveform for a binary 
system \cite{2007.book.....M} 
\begin{eqnarray}\label{eq:26}
h_+(t)= A\left(1 + \cos^2\iota \right)\cos\Phi(t)  \,, \nonumber\\
h_\times(t)= 2A\cos\iota\sin\Phi(t)  \,. 
\end{eqnarray}
Here $A=2(GM_{c}/c^2)^{5/3}(\pi f/c)^{2/3}/D_L$  is the GW 
overall amplitude, $M_c = 0.33 ~M_{\odot}$ (for component masses 
of $0.5 ~M_{\odot}$ and $0.25 ~M_{\odot}$) is the chirp mass, 
$D_L = 0.5$ kpc is the luminosity distance, 
$\iota = \pi/6$ is the inclination angle between the line of sight 
and the binary orbital axis, ${\Phi \left( t \right)}$ is the GW phase. 
For a space-based detector moving about the Sun, 
the GW phase can be given by \cite{2004PhRvD..69h2003R} 
\begin{equation}\label{eq:27}
\Phi (t)=2 \pi f t+2\pi f (R/c)\cos\theta\cos(2\pi f_{\rm{m}} t -\phi) + \varphi _0 \,,
\end{equation}
where $\varphi _0$ is the initial GW phase at the start of observation, 
$f=6.22 ~\rm{mHz}$ is the GW frequency of the reference source, 
$R=1$ AU, 
and $(\theta,\phi)$ are the ecliptic coordinates of the source. 
$h_{+,\times}$ are calculated based on the 
physical parameters of the source in \cite{2041-8205-711-2-L138}. 
Here we ignore the evolution of $f$ due to the GW emission 
from this system. The length of the simulation is one year.

\begin{figure}[htb]
\centering 
\includegraphics[width=1.07\textwidth]{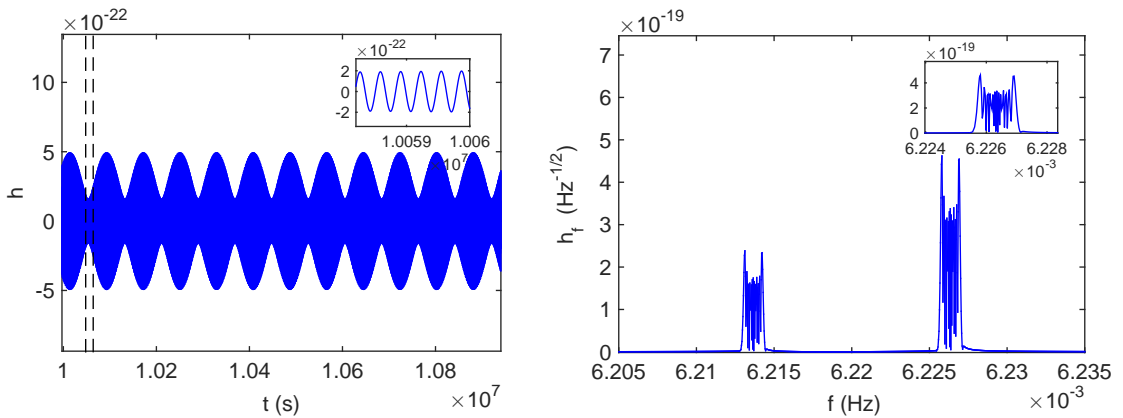}
\caption{\label{fig:4} Left panel: Clip of the one-year 
noiseless time series of the strain output of a Michelson 
interferometer for GWs from the reference source. 
The inset zooms in of the data segment within the double dash lines. 
Right panel: ASD of the time series. The inset zooms in the 
right horn. } 
\end{figure}

Shown in the left panel of Fig.~\ref{fig:4} is the strain output data for 
about 11 days from one of the Michelson interferometers. 
The inset presents the sinusoidal signal of several cycles 
within the double dash lines. One can see that 
the overall amplitude (envelope) of the signal is modulated 
by the antenna pattern being swept across the sky. 
The amplitude modulation frequency is $4f_{\rm{sc}}\approx (0.9~\rm{day})^{-1}$, 
which follows the quadrupole characteristics of 
$F_i^{+,\times}$ in Eq.~\ref{eq:apf2}. 

The phase of the signal is also modulated. This modulation 
is mainly caused by the differing time dependency of $F_i^{+,\times}$ which 
tunes the relative contribution of the two GW polarizations. 
In addition, phase modulation can also arise from the binary orbit 
axis precession, in which case the time-varying inclination 
angel $\iota$ alters the relative contribution. 
However, the orbital precession in general is negligible for 
white-dwarf binaries, therefore we do not consider this effect in 
the simulation.

Given in the right panel is the ASD $h_{f}$
of the signal. It displays two horns, of which the centers are 
separated by about $1.3 \times10^{-5}~\rm{Hz}$. 
This separation approximately equals to the modulation 
frequency of $4f_{\rm{sc}}$ due to the antenna pattern functions. 
In other words, one can regard 
the signal amplitude modulation as a beat of two sinusoidal signals 
represented by the individual horns. 
The inset enlarges the horn on the right. The broadening of individual horn mainly results from 
the Doppler modulation of the detector's motion around the Sun. 
From Eq.~\ref{eq:27}, we find that, for this case, $\Delta f/f \leq 2\pi f_{\rm{m}} (R/c)\approx 10^{-4}$, 
which makes $\Delta f \sim 6 \times10^{-7}~\rm{Hz}$.

%
%
%
%
%

\section{Conclusion and Discussions}\label{sec:sum}

TianQin is a space-based laser interferometric gravitational wave 
detector aimed at detecting GWs in the low frequencies. 
It is formed by three drag-free spacecrafts orbiting around the Earth 
in an equilateral triangular constellation. The distance between a 
spacecraft and the center of the Earth is approximately $10^5$ km. 
In this paper we provided the approximated analytic expressions for 
the coordinates of spacecraft's orbit in the heliocentric-ecliptic 
coordinate system. These expressions are sufficiently accurate and 
computationally efficient for the purpose of studying 
the scientific potential that can be enabled by TianQin as well as 
demonstrating and advancing the effectiveness of the data 
analysis techniques for various classes of GWs, such as compact 
white-dwarf binaries in the Galaxy, supermassive black hole binaries and 
extreme-mass-ratio insprirals (EMRIs), in a manner that is similar to the 
mock LISA data challenge \cite{2007CQGra..24S.551A}.

Based on this analytic orbit we calculated the detector response 
of a single Michelson interferometer to plane GWs in arbitrary 
waveform.  An example strain output for GWs from the 
reference white-dwarf binary is given which can be taken as 
the input of subsequent data analysis pipelines. Further investigations 
on simulating TDI data combinations and the associated 
sensitivities are currently in progress along with the development 
of the data analysis techniques that must content with it.

So far, the expression of the orbit does not include the perturbations from 
high order moments of the Earth, the Sun, the moon, planets, 
large asteroids, etc. The refinement of the coordinates and the numerical 
optimization of the orbit using numerical ephemerides 
\cite{2013CQGra..30f5011W} to reduce the arm length variation 
and range rate for TianQin will be subjected to our future work. 
One approach of refining coordinates is to replace the initially 
selected values of the Keplerian orbit elements and the other variables 
used in Eq.~\ref{eq:heliellip} by the time-dependent values that are 
(perhaps piecewise) fitted from the numerical integration of the orbit. 
Initial investigation shows that, for example, $\theta_s$ and $\phi_s$ 
display both secular and periodic behaviors that represent 
the wobbling of the detector plane around the 
initial direction within several degrees during the mission lifetime. 
$\boldsymbol{\varepsilon}_n(t)$ in Eq.~\ref{eq:sum} will contain the residual 
coordinates after the fit. 
The alternative approach, which is through a brutal-force, is to use 
the numerical orbit directly. For this case, one needs 
to interpolate the numerical orbit with Legendre polynomials. 
This is owning to that the 
typical time step size adopted in the numerical 
integration of the orbit  ($>10^{-3}$ day) 
is much larger than the sampling time of the 
detector output ($<1$ sec).  Due to computational cost, 
this approach can be considered in a more matured stage of the 
system simulation when the highest accuracy is required. 
For the numerical integration of the orbit, the analytic coordinates 
can be used as a guidance for setting the initial values of spacecraft 
position and velocity. 
With the more accurate orbit, we can refine the simulations of the 
response and detector output presented in this paper.

The current work serves as our first step towards building a 
system-wide end-to-end model for TianQin. 
Work to integrate realistic noise components rooted from 
different subsystems, interplanetary and relativistic effects on 
the optical path, etc. will be subjected to our future study.


\section{Acknowledgements} 
The authors thank the anonymous referees for helpful comments and suggestions.
We would like to thank Prof. Soumya Mohanty 
for helpful comments and suggestions. 
Y.W. is supported by the National Natural Science Foundation of 
China under grants 91636111, 11690021 and 11503007. 
Y.M.H. acknowledges the support from the National Natural Science 
Foundation of China under grant 11703098.

\section{References}

\appendix
\section{The coordinates of the TianQin spacecraft}\label{sec:coords} 

\begin{figure}[!h]
\centering
\includegraphics[width=0.7\textwidth]{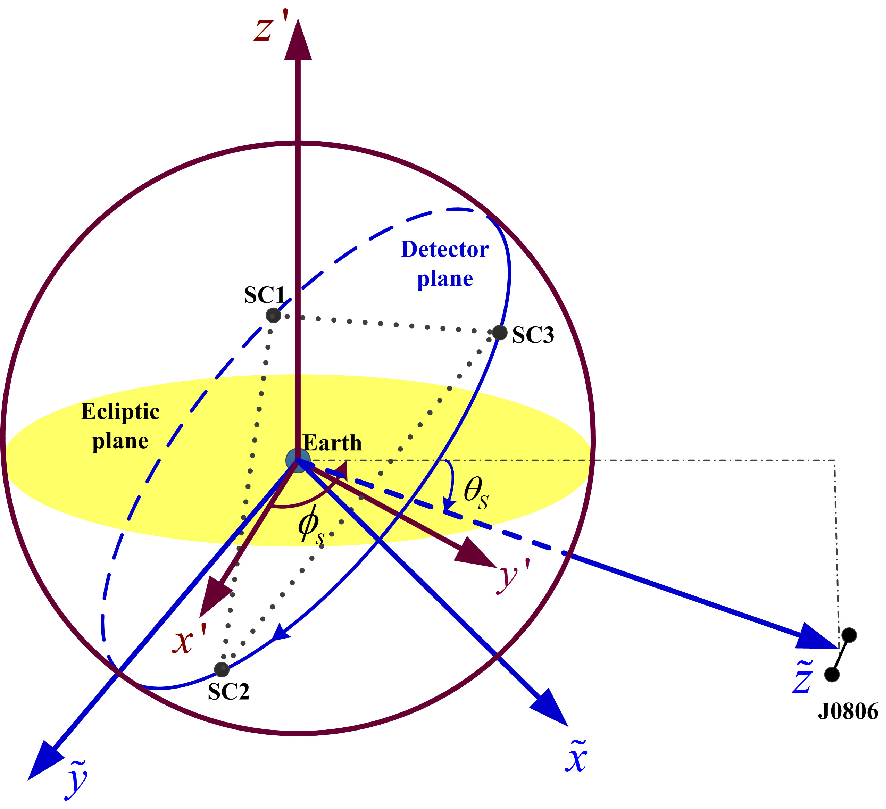}
\caption{\label{fig:2} Schematic of the detector coordinate system 
$\lbrace \tilde{x},\tilde{y},\tilde{z} \rbrace$ and the geocentric-ecliptic 
coordinate system $\lbrace x',y',z' \rbrace$. 
$\tilde{x}$ and $\tilde{y}$ axes point toward the descending node and the lowest point of the detector plane relative to the ecliptic plane, respectively. $\tilde{z}$ axis points toward the orbital angular momentum of spacecraft which is along the direction of the reference source $(\phi_s,\theta_s)$. $\lbrace x',y',z' \rbrace$ are respectively parallel to the $\lbrace x,y,z \rbrace$  in the heliocentric-ecliptic coordinate system shown in Fig.~\ref{fig:1}. Dot lines represent the optical links between each pair of spacecrafts. }
\end{figure}

We start with the Cartesian coordinates for the guiding center of 
the spacecraft constellation, i.e., the geocenter, in the 
heliocentric-ecliptic coordinate system, 
\begin{equation}\label{eq:1}
\begin{array}{l}
X = r\cos \gamma  \,,  \\
Y = r\sin \gamma  \,,  \\
Z = 0  \,,
\end{array}
\end{equation}
where $\gamma$ is the true anomaly, $r$ is the Keplerian radius
\begin{equation}\label{eq:2}
r = \frac{{R(1 - {e^2})}}{{1 + e\cos \gamma }}  \,.
\end{equation}
Here $R = 1~\rm {AU}$,
$e = 0.0167$ is the eccentricity of the Earth orbit.
The relation between $\gamma$ and the eccentric anomaly $\psi$ is
\begin{equation}\label{eq:3}
\tan \frac{\gamma }{2} = \sqrt {\frac{{1 + e}}{{1 - e}}} \tan
\frac{\psi }{2}  \,,
\end{equation}
and the relation between $\psi$ and the mean anomaly $M$ is 
\begin{equation}\label{eq:4}
\psi - e\sin \psi = M = 2\pi f_{\rm{m}} (t - \tau ).
\end{equation}
In the second equality, $M$ is expressed in terms of 
the geocenter orbit modulation frequency 
$f_{\rm{m}}=1/\rm{(one~sidereal~year)} = 3.14 \times 10^{-8}~\rm{Hz}$ 
and the passing time of the perihelion $\tau$.  In practice, it is 
more convenient to express $\psi$ by the mean orbital 
ecliptic longitude $\alpha(t)$, which is 
\begin{equation}\label{eq:5}
\psi - e\sin \psi = {\alpha }(t) - \beta \,,
\end{equation}
where $\alpha (t) = 2\pi f_{\rm{m}}t + \kappa_0$, 
$\kappa_0$ is the mean 
ecliptic longitude measured from the vernal equinox at $t=0$, 
and $\beta$ is the angle measured from the vernal equinox 
to the perihelion (see Fig.~\ref{fig:1}). Eq.~\ref{eq:5} is a transcendental equation 
which is usually solved by iterative method. 
Given $e \ll 1$, $\psi$ can be iteratively expended up to any 
order of $e$, the expression to the second order is given by 
\begin{equation}\label{eq:6}
\psi = \alpha - \beta+ e\sin (\alpha-\beta) + {e^2} \cos (\alpha-\beta) \sin(\alpha-\beta) + {\rm
O}({e^3})  \,.
\end{equation}
Next, inserting Eq.~\ref{eq:6} into Eq.~\ref{eq:3},  we find the 
expression of $\gamma $ in terms of $\alpha-\beta$ up to the 
second order of $e$ 
\begin{equation}\label{eq:7}
\gamma = \alpha - \beta + 2e\sin (\alpha- \beta) + \frac{5}{2}{e^2}\cos (\alpha- \beta)
\sin(\alpha- \beta) + {\rm O}(e^3)  \,.
\end{equation}
%
The coordinates of the geocenter can be obtained by substituting 
Eq.~\ref{eq:2} and Eq.~\ref{eq:7} into Eq.~\ref{eq:1}, which are 
\begin{eqnarray}\label{eq:8}
{X}(t) = R\cos(\alpha-\beta)+\frac{1}{2}e R\big(\cos2(\alpha-\beta)-3\big)
-\frac{3}{2}e^2R\cos(\alpha-\beta)\sin^2(\alpha-\beta)+ {\rm O}({e^3}) \,,  \nonumber\\
{Y}(t) = R\sin(\alpha-\beta)+\frac{1}{2}e R\sin2(\alpha-\beta)+\frac{1}{4}e^2 R\sin(\alpha\!-\beta)\big(3\cos2(\alpha-\beta)-1\big)+{\rm O}({e^3}) \,, \nonumber\\
{Z}(t) = 0  \,.
\end{eqnarray}

In the detector plane spanned by $\tilde{x}$ and $\tilde{y}$ axes 
shown in Fig.~\ref{fig:2}, each spacecraft revolves around the 
geocenter in a Keplerian orbit. 
We follow the same procedure as above to obtain its analytic 
coordinates in the detector coordinate system 
$\lbrace \tilde{x},\tilde{y},\tilde{z}\rbrace$. For the $n$-th 
$(n =1,2,3)$ spacecraft 
%
\begin{eqnarray}\label{eq:9}
\tilde{x}_n(t) &&=R_1\cos(\alpha_n-\beta')+\frac{1}{2}{e_1} {R_1}\big(\cos2(\alpha_n-\beta')-3\big)-\frac{3}{2}{e_1^2}R_1\cos(\alpha_n-\beta'){\sin^2}(\alpha_n-\beta') +{\rm O}({e_1^3})  \,, \nonumber\\
\tilde{y}_n(t)&&=R_1\sin(\alpha_n-\beta')+\frac{1}{2}{e_1}{R_1}\sin2(\alpha_n-\beta')+\frac{1}{4}{e_1^2}R_1\sin(\alpha_n-\beta')\big(3\cos2(\alpha_n-\beta')-1\big)+{\rm O}({e_1^3}) \,, \nonumber\\
\tilde{z}_n(t) &&= 0 \,.
\end{eqnarray}
Here $e_1$ is the eccentricity and $R_1=1.0\times 10^{5} ~\rm{km}$ 
is the semi-major axis of the spacecraft orbit. 
The spacecraft orbit phase $\alpha _n$ is 
\begin{equation}\label{eq:10}
{\alpha _n}(t) = 2\pi {f_{\rm{sc}}}t + {\kappa _n} \,,
\end{equation}
where ${\kappa _n} = \frac{2\pi}{3}(n - 1) + \lambda$, 
$\lambda$ is the initial orbit phase of the first $(n=1)$ spacecraft 
measured from $\tilde{x}$ axis. 
$f_{\rm{sc}} \approx 1/(3.65~\rm{day})$ is the modulation frequency due to 
the rotation of the detector around the guiding center. $\beta'$ is 
the angle measured from the $\tilde{x}$ axis to the perigee of the 
spacecraft orbit.

According to the relation illustrated in Fig.~\ref{fig:2}, we translate 
the coordinates $\lbrace \tilde{x}_n,\tilde{y}_n,\tilde{z}_n\rbrace$ in 
the detector coordinate system into the geocentric-ecliptic coordinate system. 
This can be done by applying two rotation matrices as below 
\begin{equation}\label{eq:12}
\left( {\begin{array}{*{20}{c}}
{x_n^{'}}\\
{y_n^{'}}\\
{z_n^{'}}\\
\end{array}} \right) = \left( {\begin{array}{*{20}{c}}
{\sin\phi_s}&{\cos\phi_s}&0\\
{-\cos\phi_s}&{\sin\phi_s}&0\\
0&0&1
\end{array}} \right)\left( {\begin{array}{*{20}{c}}
1&0&0\\
0&{\sin \theta_s }&{\cos \theta_s }\\
0&{ - \cos\theta_s}&{\sin \theta_s}
\end{array}} \right)\left( {\begin{array}{*{20}{c}}
{{\tilde x_n}}\\
{{\tilde y_n}}\\
{{\tilde z_n}}
\end{array}} \right).
\end{equation}
Here $\phi_{s}=120.5^{\circ}$ and $\theta_{s}=- 4.7^{\circ}$ 
are the ecliptic longitude and latitude of the reference source J0806.3+1527, 
respectively. Explicitly, the coordinates of the $n$-th spacecraft in the 
geocentric-ecliptic coordinate system can be written as
\begin{eqnarray}\label{eq:14}
  x'_n(t)&&=R_1\big(\cos\phi_s\sin\theta_s\sin(\alpha_n-\beta')+\cos(\alpha_n-\beta')\sin\phi_s\big)+R_1{e_1}\Big[\frac{1}{2}\big(\cos2(\alpha_n-\beta')-3\big)\sin\phi_s\nonumber\\
  &&+\cos(\alpha_n-\beta')\cos\phi_s\sin\theta_s\sin(\alpha_n-\beta')\Big]+\frac{e^2_1}{4}R_1\sin(\alpha_n-\beta')\Big[\big(3\cos2(\alpha_n-\beta')-1\big)\nonumber\\
  &&\times\cos\phi_s\sin\theta_s-6\cos(\alpha_n-\beta')\sin(\alpha_n-\beta')\sin\phi_s\Big]+{\rm O}({e^3_1}) \,, \nonumber\\
  y'_n(t)&&=R_1\big(\sin\phi_s\sin\theta_s\sin(\alpha_n-\beta')-\cos(\alpha_n-\beta')\cos\phi_s\big)+R_1{e_1}\Big[-\frac{1}{2}\big(\cos2(\alpha_n-\beta')-3\big)\cos\phi_s\nonumber\\
  &&+\cos(\alpha_n-\beta')\sin\phi_s\sin\theta_s\sin(\alpha_n-\beta')\Big]+\frac{e^2_1}{4}R_1\sin(\alpha_n-\beta')\Big[\big(3\cos2(\alpha_n-\beta')-1\big)\nonumber\\
  &&\times\sin\phi_s\sin\theta_s+6\cos(\alpha_n-\beta')\sin(\alpha_n-\beta')\cos\phi_s\Big]+{\rm O}({e^3_1}) \,, \nonumber\\
  z'_n(t)&&=-R_1\sin(\alpha_n-\beta')\cos\theta_s-R_1{e_1}\cos(\alpha_n-\beta')\sin(\alpha_n-\beta')\cos\theta_s\nonumber\\
  &&-\frac{1}{4}{e^2_1}R_1\big(3\cos2(\alpha_n-\beta')-1\big)\sin(\alpha_n-\beta')\cos\theta_s+{\rm O}({e^3_1}) \,.
\end{eqnarray}

Finally, the coordinates of the spacecraft in Eq.~\ref{eq:heliellip} can be 
easily found by adding Eq.~\ref{eq:8} and Eq.~\ref{eq:14} together. 
Note that we have only kept up to the second order of $e$ and $e_1$. 
Higher order terms can be straightforwardly obtained.

\end{document}